\documentclass[a4paper, 12pt]{article}
\usepackage[breaklinks=true,colorlinks,urlcolor=blue,linkcolor=blue,citecolor=blue]{hyperref}
\usepackage{amsmath,amssymb,amsfonts}
\usepackage{subfigure}
\usepackage{graphicx}
\usepackage{textcomp}
\usepackage{xcolor}
\usepackage{comment}
\usepackage{multirow}
\usepackage{ifthen}
\usepackage{hyperref}
\usepackage{allrunes}
\usepackage{authblk}

\newcommand{\Perthro}{\text{\textarc{p}}}    
\newcommand{\smallerfont}[1]{{\fontsize{10pt}{10pt}\selectfont#1}}
\DeclareMathAlphabet{\mathpzc}{OT1}{pzc}{m}{it}

\begin{document}

\title{On the Behaviour of Pulsed Qubits and their Application to Feed Forward Networks}

\author{Matheus Moraes Hammes}%\email{mmoraeshammes@deakin.edu.au}
\author{Antonio Robles-Kelly}%\email{antonio.robles-kelly@deakin.edu.au}

\affil{Deakin University, Waurn Ponds, VIC 3216, Australia}
%\affil[2]{Defence Science and Technology Group, Edinburgh, SA 5111, Australia}

%\keywords{Pulsed Qubits, Feedforward Networks, Sine Activations}
\date{}
\maketitle

\begin{abstract}
	In the last two decades, the combination of machine learning and quantum computing has been an ever-growing topic of interest but, to this date, the limitations of quantum computing hardware have somewhat restricted the use of complex multi-qubit operations for machine learning. In this paper, we capitalize on the cyclical nature of quantum state probabilities observed on pulsed qubits to propose a single-qubit feed forward block whose architecture allows for classical parameters to be used in a way similar to classical neural networks. To do this, we modulate the pulses exciting qubits to induce superimposed rotations around the Bloch Sphere. The approach presented here has the advantage of employing a single qubit per block. Thus, it is linear with respect to the number of blocks, not polynomial with respect to the number of neurons as opposed to the majority of methods elsewhere. Further, since it employs classical parameters, a large number of iterations and updates at training can be effected without dwelling on coherence times and the gradients can be reused and stored if necessary. We also show how an analogy can be drawn to neural networks using sine-squared activation functions and illustrate how the feed-forward block presented here may be used and implemented on pulse-enabled quantum computers.  
\end{abstract}

\section{Introduction}
Quantum computing employs quantum-mechanical phenomena to perform computations. These computing algorithms often exhibit significant increases in efficiency, in some cases exponentially, compared to their classical counterparts. As a result, since the 1970s, quantum computing has been an active area of research, and, now, as we are approaching the end of Moore's Law \cite{theis2017end}, the quest for quantum supremacy has intensified. 

Indeed, the application of quantum-mechanical phenomena to computation often yields significant increases in efficiency on quantum computers as compared to their classical counterparts. Such potential has been proven by the Shor's prime factorization \cite{ref_2} and Grover's algorithms \cite{Grover96}. Recall that Shor's algorithm is used for complex cryptography algorithms such as RSA \cite{ref_3}, whereas Grover's method is employed for search on unstructured lists. Although these methods have shown the potential computational advantages of quantum computers over classical ones, the current state of quantum hardware is still rather limited, with the largest count of qubits in a single computer being of 127 as of 2021 \cite{yang2021quantum}.

Moreover, available quantum computers, abiding by the fundamental concept of quantum information\cite{ref_5}, use reversible gates to measure and change the state of qubits. These qubits are governed by quantum mechanics, thus the effects of entanglement, interference, and superposition are manifested in their spin, polarization and energy levels \cite{ref_5}. Thus, there have been numerous research studies aiming to investigate the use of quantum computers for machine learning \cite{Lewenstein1994,Lloyd2014,Biamonte2017}. Further, in \cite{Biamonte2017}, the authors suggest quantum computers may be an ideal platform for the implementation of artificial neural networks (ANNs). This stems from the notion that the computational power of quantum computing in terms of efficiency and effectiveness combined with traditional artificial neural networks has the promise to outperform classical neural networks \cite{Jeswal2019}.

%This can be viewed as the notion that was proposed in %"A Fast Quantum Mechanical Algorithm for Database Search"
%\cite{Grover96} when they attempted to emulate quantum computation on classical computers to perform search quadratically faster than its classical equivalent in an unordered dataset.  

Nonetheless, the incorporation of quantum computation into ANNs is still an open and challenging research trend \cite{Chen2017}. This is compelled by the notion that, even theoretically, most current research in the area is focused on adapting classical neural networks to a quantum system as opposed to developing quantum neural network theory from the ground up \cite{Schuld2014}. The most prominent attempts to work within the fundamentals of quantum theory have been made as related to the quantum perceptron \cite{2014_qpercep} and the qubit neural network \cite{Kouda2005}. Moreover, Schuld et al. \cite{Schuld2014} point out how the majority of the existing approaches do not meet the requirements of what can reasonably be defined as quantum ANNs. In fact, it can be argued that the single biggest challenge for a quantum ANN is that quantum mechanics are linear, but artificial neural networks require non-linearities \cite{Cybenko1989}.

This limitation due to the linear nature of quantum mechanics was also pointed out in \cite{cao2017quantum}, where the authors propose a quantum neuron to overcome this limitation by making use of a circuit composed of two qubits. Torrontegui et al. \cite{torrontegui:2019} propose a quantum perceptron with a sigmoid activation using a qubit and a single adiabatic passage in a model of interacting spins. In \cite{farhi2018classification}, the authors present a network that employs circuits in a sequence of two qubits. Beer et al. \cite{beer2020training} treat the quantum perceptron as an arbitrary unitary operator with $m$ input and $n$ output qubits. In \cite{killoran:2019} the authors use the continuous-variable model, where quantum information is encoded not in qubits, but in the quantum states of a field, such as the electromagnetic one. Pechal et al. \cite{pechal:2021} present a superconducting qubit implementation of an adiabatic controlled gate, which generalizes the action of a classical perceptron. In \cite{arthur:2022}, the authors propose a hybrid quantum-classical neural network architecture making use of several two-qubit entangled gates.

Maybe the work that is closer to the one presented here is that in \cite{schuld:2020}, where the authors present a circuit-centric classifier that employs either single or two-qubit quantum gates. This is a low-depth circuit \cite{verdon2019quantum} which can be trained using a hybrid quantum-classical \cite{McClean2016a} gradient descent scheme. Despite some similarities, there are notable differences with respect to the method presented here. Firstly, while both employ trainable, classical parameters, the method in \cite{schuld:2020} is such that the number of learnable parameters is poly-logarithmic with respect to the input dimension. In contrast, the circuit presented here does not have such constraint, showing a behaviour much more in-line with classical neural networks. Moreover, the method in  \cite{schuld:2020} encodes the input feature vectors into the amplitudes of a quantum system. Rather, our approach employs a pulse modulator to excite the qubit. This is important since, as noted in  \cite{schuld:2020}, quantum methods that are polynomial in the number $n$ of qubits can perform computations in 2$^n$ amplitudes. This does not apply to the circuit presented here. 

Here, we note that the underlying physics in pulse-enabled quantum computers offers the opportunity to exploit the cyclical nature of pulsed qubits to implement feed fodward networks. Thus, we propose a single-qubit feed forward block whose architecture allows for classical parameters to be used in a manner akin to classical neural networks. We do this by using the rotations of pulsed qubits on the Bloch sphere to show how such a block would be reminiscent of a single-layer neural network comprised of neurons with sine-squared activation functions. In contrast with other methods elsewhere, the number of qubits is not a function of the number of neurons in the network, but rather is linear in the number of blocks in the resulting circuit.  Moreover, keeping the parameters classical has the further advantage that a large number of iterations and updates can be effected without dwelling on coherence times, whereby the gradients can be reused and stored if necessary and standard libraries can be used for loss minimization. 

%Continue the literature review on QNNs here

% Do the closing paragraph of the intro at the end

\section{Quantum Pulses}

As mentioned earlier, quantum computers are an emerging technology. Operations in quantum systems arise from precise pulses stimulating qubits to change their state. The importance of pulse-level programming resides in its capacity to allow users full control of the quantum transformation via programmable pulses and their scheduling. To our knowledge, however, pulses have only been used to calibrate quantum gates for better precision and efficiency by changing the qubit states \cite{ref_10}. This improved precision when manipulating qubits is needed to achieve quantum supremacy, as researchers try to control the uncertainty of superposed states to reduce the quantum error.

Quantum pulse level programming utilises resonance to modify the state of a qubit. A drive pulse will excite a qubit when it resonates with its natural frequency, rotating it around the Bloch Sphere. This is expected since the qubit frequency is the difference between the ground and excited states, i.e. $\vert 0\rangle$ and $\vert 1\rangle$ in ``Bra-ket'' notation.

For the purposes of this paper we will focus on Gaussian pulses and, where relevant, will use Bra-Ket notation. This is as Bra-Ket notation captures the pairing of a linear function and complex vector in a Hilbert space. This is done as follows  $$\vert  0 \rangle  = \left[ \begin{array}{l}1\\0\end{array} \right] \text{\hspace{0.5cm}and\hspace{0.5cm}}\vert  1 \rangle  = \left[ \begin{array}{l} 0\\ 1 \end{array} \right]$$.

\subsection{The Bloch Sphere}
%\paragraph{}

Recall that a single qubit, much like their classical computing counterparts, is a two-state system. These states are $\vert 0\rangle$ and $\vert 1\rangle$. The correspondence of quantum states and binary values is the base for quantum information theory, where information is represented by quantum states and, with the use of quantum mechanics, quantum computations can be performed accordingly. Qubits are commonly defined by the following equation: 
\begin{equation}
\vert \psi\rangle=\alpha\vert 0\rangle+\beta\vert 1\rangle,
\label{eq:01}
\end{equation}
with $\alpha, \beta \in$ $\mathbb{C}$ and $\vert 0\rangle, \vert 1\rangle$ in the Hilbert space $ H^2$, where $\vert \alpha\vert $ and $\vert \beta\vert $ are probability amplitudes. 

Thus, measuring a qubit makes $\vert \psi\rangle$ collapse into either $\vert 0\rangle$ or $\vert 1\rangle$with probabilities $\vert \alpha\vert ^2$ or $\vert \beta\vert ^2$ such that
\begin{equation}
\vert \alpha\vert ^2 + \vert \beta\vert ^2 = 1
\label{eq:02}
\end{equation} 

The Bloch sphere, hence, represents geometrically the state of a quantum system in a two-dimensional Euclidean space. By arbitrarily choosing a pair of antipodal points as the two basis vectors $\vert 0\rangle$ and $\vert 1\rangle$ of a quantum system, it is possible to visualise the ``pure'' state of a qubit. The defining equation for representing a state in a Bloch sphere is derived from the general state equation given by
\begin{equation}
\vert \psi \rangle =\alpha \vert 0\rangle +\beta \vert 1\rangle
\label{eq:03}
\end{equation} 
where $\alpha$ and $\beta$ describe the probability of the qubit collapsing into either of its two possible states when measured.    % Note these are both complex numbers. 

After normalising the qubit state, applying the trigonometric identity and removing the global phase term, we note that, for measuring a single qubit, its state $\vert \psi \rangle$ can be expressed as follows
\begin{equation}
\vert \psi \rangle = \cos{\tfrac{\theta}{2}}\vert 0\rangle + \exp\big({\mathbf{i}\phi}\sin{\tfrac{\theta}{2}}\vert 1\rangle\big)
\label{eq:04}
\end{equation}
such that $\theta ,\phi \in \mathbb{R}$ are the angles on the Bloch sphere.

\begin{figure}[t!]
	\centering
	\includegraphics[width=0.45\columnwidth]{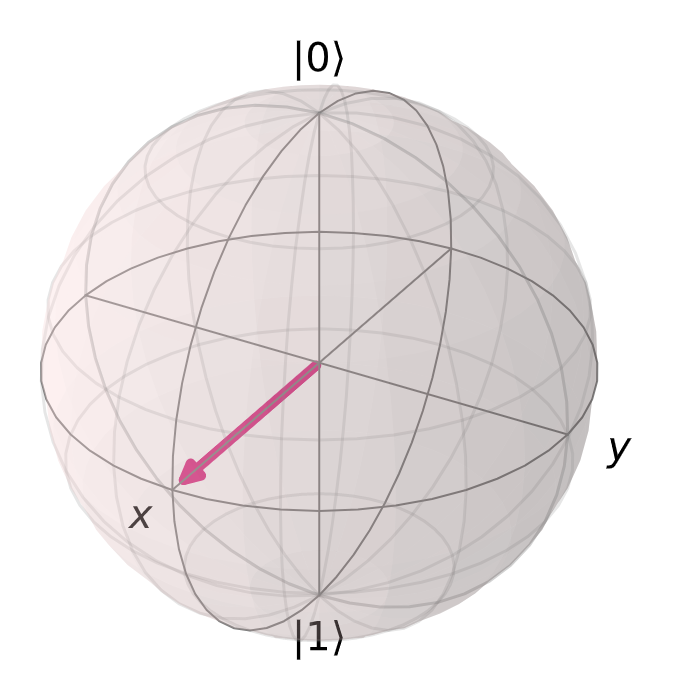}
	\includegraphics[width=0.45\columnwidth]{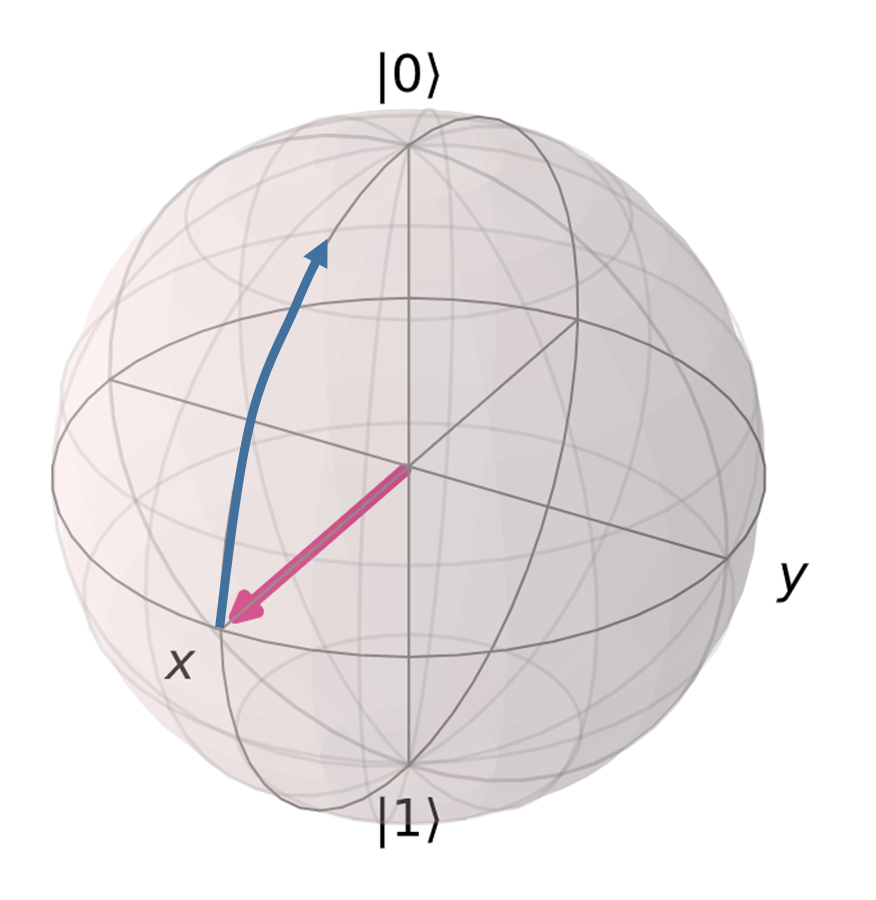}
	\caption{Left-hand panel: A Bloch sphere where the state of a qubit is such that it has equal probability of falling into either of its two states $\vert 0\rangle$ or $\vert 1\rangle$; Right-hand panel: A Bloch sphere where superposition has been applied to the qubit in the left-hand panel, rotating the qubit towards the $\vert 0\rangle$ state.}
	\label{fig:bloch_superposition}
\end{figure}

Thus, the probability of a qubit collapsing into a $\vert 0\rangle$ or $\vert 1\rangle$ state when measured is indicated by the angle $\theta$. The closer to a pole, the more likely the qubit is to collapse into the state corresponding to that particular pole. A $90^o$ angle thus represents the point where the qubit has an equal probability of collapsing into either state. We illustrate this in the left-hand panel of Figure \ref{fig:bloch_superposition}, where we show the representation of a qubit that has equal odds of falling into either state. Futher, the cyclic oscillation of states in the presence of a driving pulse in a quantum system is called a Rabi Cycle. A pulse that drives a qubit from a $\vert 0\rangle$ to $\vert 1\rangle$ is called a $\pi$-pulse, being equivalent to a 180\textdegree \space rotation on the Bloch sphere. It is worth noting in passing that this is analogous to an X180 qubit flipping gate \cite{ref_11}.

%\subsection{$\pi$ Pulses}

Recall that defining a $\pi$-pulse in the context of a qubit requires two experiments to be conducted. Firstly, a frequency sweep. Secondly, a Rabi experiment. The frequency sweep experiment tests the resonance frequency of the qubit by sweeping over a range of pulses with different frequencies, measuring the energy absorption of the system for each of these. The resonant frequency of the qubit corresponds to the peak of the absorption curve.   
%We illustrate this in Figure \ref{fig:f_sweep}, where we show a frequency sweep experiment, where the magnitude of excitation of a qubit is shown as a function of $\omega$ for varying frequencies. In the plot, each of the points corresponds to a measurement of the excitations of the qubit. 
Once the resonant frequency of the target qubit is found, the amplitude of the $\pi$-pulse can be defined by varying its amplitude $\alpha$ so as to make the qubit oscillate around the Bloch sphere. The amplitude required to shift a qubit from its ground to its excited state is then given by half the period corresponding to the resonant frequency.

\begin{figure}[b!]
	\centering
	\includegraphics[width=.47\columnwidth]{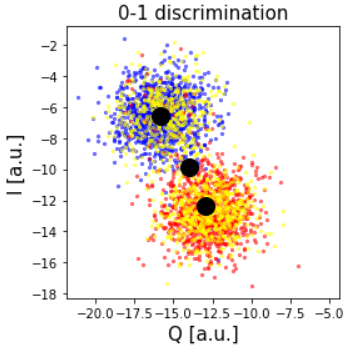}
	\includegraphics[width=.47\columnwidth]{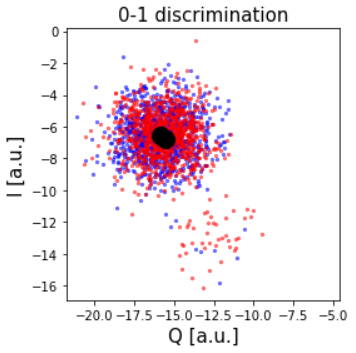}
	\caption{Left-hand panel: Half $\pi$-pulse measurements; Right-hand panel: Double $\pi$-pulse measurements showing the corresponding shift induced by the second pulse with respect to the left-hand panel.}
	\label{fig:half_disc}
\end{figure}

\subsection{Pulsing Qubits}

To further understand the qubit behaviour as related to the equations above, consider the case where the amplitude of a pulse is doubled. This will induce the qubit to move 360\textdegree \space around the Bloch sphere, returning it to its original state. On the other hand, halving the amplitude of a pulse drives a qubit to rotate 90\textdegree \space around the Bloch sphere, leaving the qubit in a symmetric superposition state. This is essentially equivalent to a Hadamard gate. This can be appreciated in Figure \ref{fig:half_disc}, where we show the qubit measurements over 128 trials as a function of the two components, real ($Q$) and imaginary ($I$), in Equation \ref{eq:02} when half and double shifts are applied to the qubits. In the plots, each of the points corresponds to a measurement of the qubit and the black markers correspond to the population mean.  In both plots, red denotes ground states and blue excited ones. On the left-hand panel we have also included superimposed states, which we plot in yellow.

As mentioned earlier, here we focus on Gaussian pulses. Being more formal, these can be defined as follows
\begin{equation}
\Upsilon\left( \omega\right) =\mathpzc{A}\exp\left( -\dfrac{\left( \omega-\gamma\right) ^{2}}{2\tau^{2}}\right)
\label{eq:05}
\end{equation}
where $\omega$ is the frequency of the pulse, $\gamma$ is the duration of the pulse, $\mathpzc{A}$ is the amplitude and $\tau$ governs its width. 

For Gaussian pulses, the resulting excitation plot is a sinusoid that shows direct correlation to the amplitude $\mathpzc{A}$ of the pulse. Note that this is consistent with the $\pi$-pulse measurments shown in Figure \ref{fig:half_disc} and can be related to the the Bloch Sphere using the real component of the second term on the right-hand side in Equation \ref{eq:04}.   
This can be express the excitation of the qubit as follows
\begin{equation}
g\left( \mathpzc{A}\right) = a\sin \left(\eta\mathpzc{A}\right)+c
\label{eq:06}
\end{equation}
where we have used the the identity $\sin\left(\eta\mathpzc{A}\right)=\cos \left(\frac{\pi}{2}-\eta\mathpzc{A}\right)$, $\eta$ and $a$ are proportionality constants and $c$ is a constant that captures the vertical shift of the sinusoid. 

From the equation above it becomes straightforward to note that the qubit state after being excited by a pulse is determined by the amplitude $\mathpzc{A}$. Thus, by using the amplitude, the pulse can be ``shifted'' around the Bloch sphere, allowing for the control of output signals through superposition. To do this, we set, without any loss of generality, $\eta$ and $a$ to unity and constrain $\mathpzc{A}\in[0,1]$.  By using the amplitude $\mathpzc{A}$ and properties of $\pi$ rotations on the Bloch sphere, we can write the probability of measuring the qubit in the state $\vert 1\rangle$ as follows  
\begin{equation}
%I\kern-0.22em 
P({\vert 1\rangle})  = \sin^2 \bigg(\frac{\pi}{2}\mathpzc{A}\bigg)
\label{eq:07}
\end{equation}

\subsection{On Superposition and Feedforward Networks}

% \subsection{Sine Networks}
%whereby $\mathpzc{A}$ can be viewed as a modulating amplitude of the pulse and, consequentially, modifying the value of the $\theta$ rotation in Equation \ref{eq:03} around the Bloch Sphere. 

Note that Equation \ref{eq:07} is somewhat reminiscent of a neuron in a network with a squared sinusoidal activation function. To see this more clearly,  let $f\left( \mathbf{x}\right)=P({\vert 1\rangle})$ and proceed to rewrite Equation \ref{eq:07} as follows 
\begin{equation}
f\left( \mathbf{x}\right) = \sin^2 \left(\langle\mathbf{w},\mathbf{x}\rangle+b\right)
\label{eq:08}
\end{equation}
where $\langle\cdot,\cdot\rangle$ denotes, as usual, the dot product, $\mathbf{x}$ is the input vector, $\mathbf{w}$ is the weight vector, $b$ corresponds to the bias, and $\sin^2(\cdot)$ is the activation function.  Moreover, note that the amplitude of the pulse $\mathpzc{A}$ can be viewed as the modulator of the rotation in Equation \ref{eq:03} around the Bloch sphere. Thus, it becomes straightforward to note that, by amplitude-modulating the pulses such that
\begin{equation}
\mathpzc{A} = \frac{2}{\pi}\left(\langle\mathbf{w},\mathbf{x}\rangle+b\right)
\label{eq:09}
\end{equation}
Equations \ref{eq:07} and \ref{eq:08} become equivalent.

Making use of Equation \ref{eq:09} and the principles of superposition on the Bloch sphere, we can now explore the use of a train of pulses instead of a single one. Note that, on the Bloch sphere, each subsequent pulse will have the effect of rotating the qubit state towards either pole. From the previous sections, it's straightforward to note that these rotations are governed by the amplitude of the pulses as applied to the qubit excitation in Equation \ref{eq:04}.

This is an important observation since, by indexing the amplitude to the pulse position on the train, one could explore the use of these amplitudes and their relationship to the weight and bias vectors in Equations \ref{eq:08} and \ref{eq:09} to obtain a behaviour reminiscent of that exhibited by the layers of a neural network with a squared-sine activation function. Thus, let $\mathpzc{T}$ be a train of $n$ pulses with amplitudes $\mathpzc{A}_i$, where $i$ is the index of the pulse in the train. The probability of a qubit being in a state $\vert 1\rangle$ for the $i^{th}$ pulse is then given by
\begin{equation}
%I\kern-0.22em 
P({\vert 1\rangle})  = \sin^2 \bigg(\frac{\pi}{2}\sum_{j=1}^i\mathpzc{A}_j\bigg)
\label{eq:10}
\end{equation}

\begin{figure}[t!]
	\centering
	%\fbox{
	\includegraphics[width=0.95\textwidth]{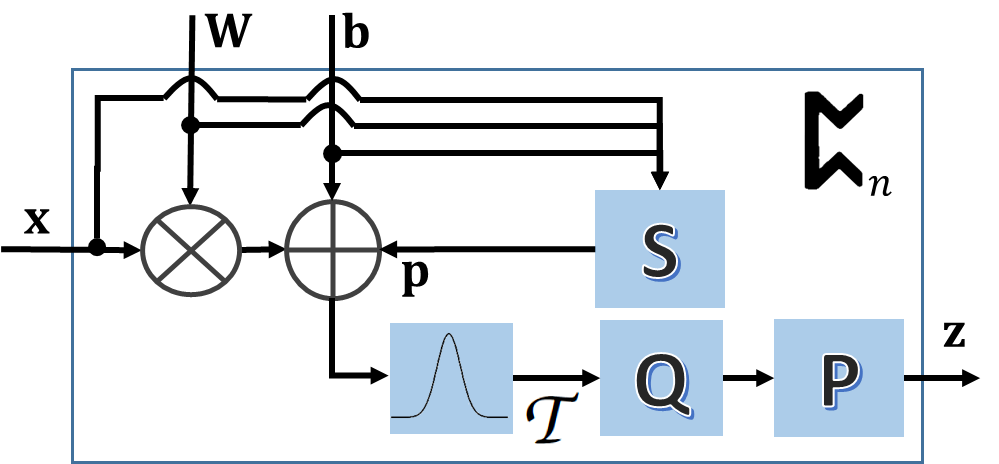}
	%}
	\caption{Conceptual diagram of the $\Perthro$ block presented here. The block takes at input $\mathbf{x}$ and, making use of the weight matrix $\mathbf{W}$ and bias vector $\mathbf{b}$ modulates a train of pulses $\mathpzc{T}$ so excite the qubit $\mathrm{Q}$ to deliver at output the vector $\mathbf{z}$ of measured probabilities.}
	\label{fig:perthro_gate}
\end{figure}

\section{Pulsed Qubits for Feed Forward Neural Blocks}

Note that the equation above is telling. This is since it points to the fact that, as a train of pulses is used to excite the qubit, its probability will rotate on the Bloch sphere such that for $n$ pulses there will be the same number of rotations. Each of these rotations corresponds to the state probabilities of the qubit. If we view Equation \ref{eq:08} as that corresponding to the output of a neuron on a neural network we could then draw an analogy of a qubit excited by a train of pulses to a neurla block comprised by a single-layer of $n$ neurons. These single blocks would then be somewhat reminiscent of layers in classical neural networks. This is important since a block would then be comprised of a single qubit. This is, the number of qubits is linear to the number of blocks, not to the number of neurons. We discuss this further in Section \ref{sct:discussion_and_behaviour}.

%, but with notable differences. 
For now, we focus our attention on the pulsed qubits and explore the analogy to a single-layer block of $n$ neurons. To do this, let $\mathbf{w}_i$ and $b_i$ be the weight vector and bias in Equation \ref{eq:08} corresponding to the amplitude in the $i^{th}$ pulse. Under the analogy above, these would also be the weights and bias of the $i^{th}$ neuron in our neural network layer. With this notation, and making use of Equation \ref{eq:09} we can write Equation \ref{eq:10} as follows
\begin{eqnarray}
\nonumber
P({\vert 1\rangle})& = &\sin^2 \bigg(\sum_{j=1}^i\big(\langle\mathbf{w}_j,\mathbf{x}\rangle+b_j\big)\bigg)\\
& = & \sin^2 \big(\langle\mathbf{w}_i,\mathbf{x}\rangle+b_i+\varrho_i\big)
\label{eq:11}
\end{eqnarray}
whereby, in the equation above, we have introduced the term 
\begin{equation}
\varrho_i = \sum_{j=1}^{i-1}\big(\langle\mathbf{w}_j,\mathbf{x}\rangle+b_j\big)
\label{eq:12}
\end{equation}
which, by definition, is set to zero for $i=1$.

From Equation \ref{eq:12} it becomes clear that $\varrho_i$ should be taken into account as a result of the superposition of states in qubits. That said,  it can be easily computed if the weights and biases are in hand. In Figure \ref{fig:perthro_gate}, we show the diagram of a quantum block, which we denote $\Perthro_n$\footnote{We have chosen the Perthro rune symbol since it shares its transliteration with $\pi$ in an analogy to the common notation used for pulsing qubits.}, aimed at obtaining a vector $\mathbf{z}$ of measured probabilities corresponding to the train of $n$ pulses $\mathpzc{T}$ being applied to the qubit $\mathrm{Q}$. 

In Figure \ref{fig:perthro_gate}, we have denoted $\mathbf{W}$ as the matrix of weights whose $i^{th}$ row corresponds to the vector $\mathbf{w}_j^T$. Similarly, we have used $\mathbf{p}$ and $\mathbf{b}$ as the vectors whose entry indexed $i$ corresponds to $\varrho_i$ and $b_i$, respectively. In Figure \ref{fig:perthro_gate}, the block $\mathrm{S}$ computes the vector $\mathbf{p}$, which is added to the bias and the product of the input and the weight matrix $\mathbf{W}$. This yields a vector of amplitudes that are passed on to the pulse generator, which applies the train of $n$ pulses $\mathpzc{T}$ to the qubit. The block $\mathrm{P}$ delivers the vector $\mathbf{z}$ of measured probabilities at output. This is such that the $i^{th}$ entry of $\mathbf{z}$ corresponds to the qubit probability $P(\vert 1\rangle)$ corresponding to the pulse indexed $i$ in $\mathpzc{T}$.

\section{Discussion}
\label{sct:discussion_and_behaviour}

\subsection{Relation to Neural Blocks}

Note that the block $\Perthro_n$ is somewhat reminiscent of a single-layer neural network block. Indeed, Artificial Neural Networks rely on activation functions to produce non-linear outputs based upon the inputs of the neuron under consideration. That said, there are notable differences. Firstly, most of the widely used functions, such as ReLU, Sigmoid, Tanh, and Leaky ReLu are monotonic. This often improves convergence, as the output of the neuron can be viewed as being correlated to the input, whereby larger weights imply a larger correlation of the input to the output and a weak correlation is reflected by smaller weights. 

Non-monotonic activation functions behave differently. Moreover, sine activation functions, due to their periodicity, can be viewed as oscillators. This is, for a large correlation of inputs, the activation function oscillates from weaker to stronger activations i.e. outputs may increase or decrease as weights increase. This oscillation is regarded as an inconvenience to general learning tasks, with the widespread belief that sine excitation functions in neural networks are harder to train and slower to converge \cite{ref_7}. 

As a result, sine neural networks are considerably under-researched and seldom used. Despite the small volume of research concerning neural networks using sinusoidal activation functions, these functions have been successfully implemented as hidden layers in neural networks - at times improving accuracy and convergence at training as compared to their sigmoidal counterparts \cite{ref_8}. In addition to performance and accuracy improvements, periodic activation functions can offer larger memory capacity in chaotic neuron models\cite{ref_9}.

More importantly, the inclusion of the term $\varrho_i$ in Equation \ref{eq:11} implies that the $i^{th}$ output of the block would depend not only on the weights and biases indexed $i$ but also on the ones for which $j< i$. This, as mentioned earlier, is due to the superposition on the Bloch sphere.  There are several ways in which  $\varrho_i$ may be interpreted. Maybe the most intuitive of these is as a cumulative phase variable on the pulse modulator. Thus, the longer the train of pulses $\mathpzc{T}$, the more terms the right-hand side of Equation \ref{eq:12} will comprise.

\subsection{Pulse Programming}

Despite being fundamental to the calibration of qubits, it was until recently that pulse-level programming was available on an open-source SDK. Indeed, recently, IBM has released Qiskit, an SDK that allows for working with quantum computers at the level of pulses, circuits and application modules. Further, Qiskit is an open-source framework dedicated to creating and experimenting with quantum programs on either simulated or experimental quantum computers, enabling developers to remotely program on quantum devices. Qiskit at first only offered unitary gates that allowed the manipulation of the state of qubits but, in a recent update, manipulating the inherent quantum physical dynamics was made possible with the addition of pulse-level programming \cite{ref_10}. Hence, we use it henceforth for the results in this paper, which were carried out on IBM's Armonk quantum computer.

As Qiskit pulse-enabled qubits have a target frequency of 5-7GHz, a frequency sweep is needed to fine-tune the pulses while accounting for variations in the system or the environment. It is worth noting that Qiskit does not allow in general more than four pulses to be used on a single train. As a result, in the results and experiments shown here, we have accounted for this by allowing the qubits to reset when necessary and accounted for the corresponding amplitude shifts accordingly. Despite these limitations, quantum pulse fidelity and qubit calibration are active areas of research. Recent papers reportedly have achieved an accuracy of over 99\% \cite{2016Ballance}, considered to be the threshold for fault-tolerant quantum computation \cite{2005Knill}.

%All our experiments 

\begin{figure}[t!]
	\centering
	\fbox{\includegraphics[width=0.9\textwidth]{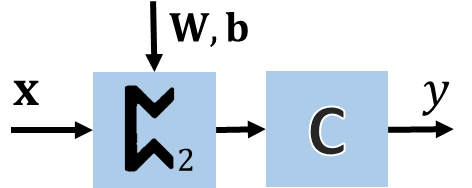}}\\
	%\vspace{0.3cm}
	%\fbox{\includegraphics[width=0.9\columnwidth]{./images/XOr_net_full.png}}
	\caption{Block diagram of the feed forward circuit used for the XOR simulation. The circuit is comprised of a single Perthro block and $\mathrm{C}$, which is a thresholding one.}
	\label{fig:XOr}
\end{figure}

% Continue here
\section{Experiments}

In this section, we show results on three tasks using the Perthro gate presented earlier. The first of these is the simulation of an XOR gate. The second one pertains to categorical classification and the final one is a regression task. To obtain the results for our approach shown throughout the section, we have made use of Qiskit pulse schedules on IBM's Armonk quantum computer to feed the single qubit in each of the $\Perthro_n$ blocks with the corresponding pulses of different amplitudes \footnote{The qubit has also been calibrated using the procedure in \url{https://qiskit.org/textbook/ch-quantum-hardware/calibrating-qubits-pulse.html}}. We have also bound the amplitude pulses making use of the drive current obtained after calibration. All the jobs were initialized and sent to the quantum computer using a personal laptop with a 2.7 GHz Dual-Core Intel i5 processor and 8 GB. This laptop was also used to process the results of each job and optimize model parameters accordingly.

The use of Qiskit also permits finding the values of the weight matrices and the bias vectors used throughout the section. To this end, we have implemented the Perthro block equations presented earlier in a simulator on a classical computer and trained it using both, the simulator and tensorflow. This hinges in the notion that we can view the Perthro block $\Perthro_n$ as an $n$-neuron single layer network, where each neuron output corresponds to Equation \ref{eq:11}. This allows for the weights to be found by minimising, via gradient descent, the corresponding loss in a manner akin to that used in the classical training of feedforward networks while performing the forward evaluation of the network using the simulator. Once the parameters are in hand, these can be used to modulate the pulses using Qiskit and tested on IBM’s Armonk quantum computer. 
%In our optimisations, we have constrained all our weight matrices and biases to be in the range $[0,1]$.

\subsection{XOR Computation}

To illustrate the behaviour of the \Perthro$_n$ block introduced earlier, we now turn our attention to reproducing an XOR gate. Note this is a classical problem in artificial neural network research. The problem is that, given two binary inputs, predict the output of an exclusive-or gate.  An XOR gate returns a false value (a zero) if the inputs are equal and true (unity) if the inputs are not. In this case, we have used this simulation in order to illustrate the behaviour of the block when used in an analogy to a two-neuron perceptron, i.e. single-layer network, with a classification stage. 

Here, we employ for our experiments a circuit comprised by  one Perthro block. In Figure \ref{fig:XOr}, we show the block diagram of our feed forward circuit, where the Perthro block $\Perthro_2$ is such that it takes at input two bits and delivers, at output two probabilities following Equation \ref{eq:11}. The  block $\mathrm{C}$ is a classification stage, which in this case, is based on two thresholds. To obtain the final XOR outputs, we use the output of the $\Perthro_2$ block, which we denote as $\mathbf{z}=[P_1(\vert 1\rangle), P_2(\vert 1\rangle)]^T$ and set the output of $\mathrm{C}$ such that the prediction of the circuit becomes
\begin{equation}
y = \begin{cases}
0 &\text{if\hspace{0.5cm}}P_1(\vert 1\rangle)\geq 0.5 \  \wedge  \ P_2(\vert 1\rangle)\leq 0.5\\
1 &\text{Otherwise}
\end{cases}
\end{equation}

\begin{figure}[t!]
	\centering
	\includegraphics[width=\textwidth]{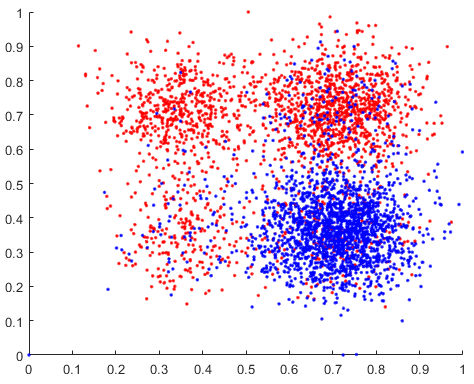}
	\caption{Scatter plot for the two qubit probabilites of $\vert 1\rangle$ delivered by the block $\Perthro_2$ in the XOR simulation circuit shown in Figure \ref{fig:XOr_results}. In the scatter plot, we have denoted the variables whose ground-truth values correspond to logical 1 in red and 0 as blue.}
	\label{fig:XOr_results}
\end{figure}

In Figure \ref{fig:XOr_results}, we show the scatter plot for the output vector $\mathbf{z}$ over 1024 trials for the block $\Perthro_2$, where the x-axis corresponds to the probability  $P_1(\vert 1\rangle)$ and the y-axis to $P_2(\vert 1\rangle)$. In the figure, we show in blue those trials for which the ground-truth, i.e. the output of an XOR gate for the input vector $\mathbf{x}$, is one. In red, we show those for which the trail should yield a zero. The scatter plot also illustrates the probabilistic nature of quantum physics, whereby the output of the block $\mathrm{C}$ when applied to the measured probabilities corresponds to the ground truth 88.77\%, i.e. 908 out of 1024 trials correspond to the ground truth after the measured probabilities have been thresholded. The thresholds were chosen to reflect the notion that $0.5$ is where the qubit has an equal chance of collapsing into either the $\vert 1\rangle$ or $\vert 0\rangle$ state.

%We now turn our attention to the effect of adding another block to the circuit in the left-hand panel of Figure \ref{fig:XOr}. In contrast with the first of the circuits under consideration here, the addition of the second block will produce a single output probability, which again, can be converted into a binary output using the threshold block $\mathrm{C}$. In our experiments,
%those trials for which the ground-truth is one and those for which the trial should yield a zero deliver well-separated distributions, with means at $0.049$ and $0.7974$ for zero and one trials, respectively. Moreover, when setting the threshold in the block $\mathrm{C}$ to $0.5$, i.e. the value where the qubit has an equal chance of falling into either state, 88.77\% of the trails correspond to the ground truth at the output.

%In the right-hand panel of Figure \ref{fig:XOr_results}, we show the histogram of output probabilities for $\Perthro_1$ block in the circuit shown in the right-hand panel of Figure \ref{fig:XOr}  over 1024 trials. In the histogram, the bars corresponding to the count of 
%those trials for which the ground truth is one in blue. In red, we show those for which the trail should yield a zero. 

\begin{table}[!b]
		\begin{center}
		{\normalsize
			\begin{tabular}{|l|l|l|l|}
			\hline
				\multicolumn{1}{|l|}{\smallerfont{Number of $\Perthro$ blocks}} & \multicolumn{1}{l|}{\smallerfont{Number of Pulses}} &  \multicolumn{1}{l|}{\smallerfont{Number of Neurons}} & \multicolumn{1}{l|}{\smallerfont{Number of Qubits}}\\% &  \multirow{2}{*}{Error} \\ 
				\hline
%				\multicolumn{1}{|l|}{\smallerfont{Number of}} & \multicolumn{1}{l|}{\smallerfont{Number}} &  \multicolumn{1}{l|}{\smallerfont{Equivalent Number}} & \multicolumn{1}{l|}{\smallerfont{Number}}\\% &  \multirow{2}{*}{Error} \\ 
%				\multicolumn{1}{|l|}{\smallerfont{$\Perthro$ blocks}} & \multicolumn{1}{l|}{\smallerfont{of Pulses}} &  \multicolumn{1}{l|}{\smallerfont{of Neurons}} & \multicolumn{1}{l|}{\smallerfont{of Qubits}} \\ 
%				 &  &  \multicolumn{1}{l|}{\smallerfont{of Neurons}} & \\ 
				% & \multicolumn{1}{l|}{blocks} & \multicolumn{1}{l|}{Pulses} &  \multicolumn{1}{l|}{of Neurons} & \multicolumn{1}{l|}{Qubits} & \multicolumn{1}{|l|}{}   & \multicolumn{1}{|l|}{} \\ 
				\hline
				\multicolumn{1}{|l|}{ 1} & \multicolumn{1}{l|}{ 2}        & \multicolumn{1}{l|}{2}      & \multicolumn{1}{l|}{1}          %& \multicolumn{1}{l|}{0.1133}        
				\\ \hline
				%\multicolumn{1}{|l|}{XOr} & \multicolumn{1}{l|}{ 2} & \multicolumn{1}{l|}{ 3}        & \multicolumn{1}{l|}{3}      & \multicolumn{1}{l|}{2}        & \multicolumn{1}{l|}{0.1649}        \\ \hline
			\end{tabular}
	}
	\caption{Statistics for the XOR circuit used in our experiments. }
	\end{center}
	\label{table:results}
\end{table}

To find the values of the weight matrix $\mathbf{W}$ and the bias vector $\mathbf{b}$ for the circuit in Figure \ref{fig:XOr_results}, we have employed the mean squared error (MSE) loss of the predictions with respect to the ground truth in a manner akin to that used in feed forward networks. Note that, despite this minimisation hinges on the equivalent of a two-neuron network, the Perthro blocks $\Perthro_2$ is built upon a single qubit, whereby the pulse modulator delivers two pulses to the qubit. These pulses account for each of the two probabilities in $\mathbf{z}$ used to produce the scatter plot shown in Figure \ref{fig:XOr_results}. This is reflected in Table \ref{table:results}, where we show the statistics for our XOR circuit.

%These outputs correspond to the rotations of the qubit on the Bloch Sphere.
\begin{figure}[b!]
	\centering
	% \fbox{\includegraphics[height=0.8in]{./images/regression_net.png}}
	\fbox{\includegraphics[width=0.9\textwidth]{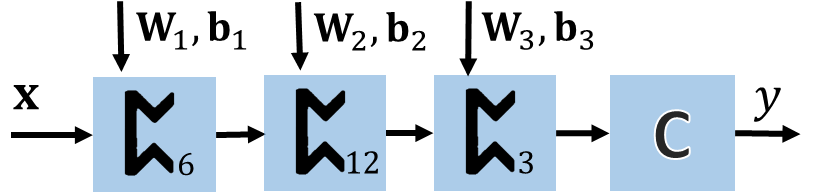}}
	\caption{Conceptual diagram of the circuit used for our classification experiments. Note each of the blocks comprises a qubit, regardless of how many pulses are used. In this case the $\mathrm{C}$ block is a classical soft-max classification stage applied after the last Perthro block $\Perthro_3$ delivers its $\mathbf{z}$ output vector of measured probabilities.}
	\label{fig:Pertho_Nets}
\end{figure}

% Continue here

\subsection{Classification}

Note the XOR experiments above are consistent with the notion that the Perthro blocks can be used as layers in neural networks. As a result, we now turn our attention to a classification task involving longer pulse trains, i.e. pulse schedules. To illustrate this, we now turn our attention to a classification problem using the Iris dataset \footnote{The Iris dataset can be accessed at \url{https://archive.ics.uci.edu/ml/datasets/iris}}.

For our experiments on the Iris dataset, we have used the circuit in Figure \ref{fig:Pertho_Nets}. In the figure, to avoid confusion, we have indexed the weight matrices and bias vectors to the block number. For the training of these weights and biases, again, we have viewed the blocks $\Perthro_6$,  $\Perthro_{12}$ and  $\Perthro_3$ as single layers of 6, 12 and 3 neurons whose activation function corresponds to Equation \ref{eq:11}. Note that, viewed in this manner, the circuit is equivalent to a 21 neuron, three-layer, feed forward network whereby the weight matrices and biases can be learnt accordingly. In the circuit, the block $\mathrm{C}$ now corresponds to a softmax classification stage and, hence, to train it, we have used the categorical cross-entropy loss. This is telling since the circuit in Figure \ref{fig:Pertho_Nets} only employs three qubits in overall.

\begin{figure}[t!]
	\centering
	\includegraphics[width=1\textwidth]{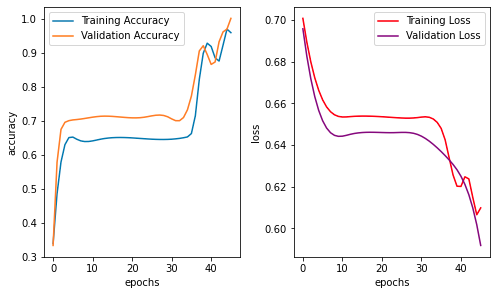}
	\caption{Training accuracy (left-hand panel) and loss (right-hand panel) as a function of iteration number for the classification circuit in Figure \ref{fig:Pertho_Nets} used in our experiments on the Iris dataset.}
	\label{fig:training_iris}
\end{figure}

In Figure \ref{fig:training_iris}, we show the training loss and categorical accuracy for the training as a function of iteration number. In our classification experiments, we performed both the simulation using Qiskit and the test on IBM's Armonk quantum computer. For these, the test accuracy and standard deviations delivered by our simulated circuit were $79.29\pm 5.04\%$, whereas IBM's computer yielded $73.33\pm 10.4\%$. Note these deliver a margin of improvement over those reported for the quantum-classical neural network architecture in \cite{arthur:2022}. Further, note that, in \cite{arthur:2022}, the authors achieved a simulated variational quantum circuit (VQC) accuracy of $81.5\pm 14.37$ and a simulated hybrid neural network (HNN) accuracy of $89.88\pm 4.24\%$ using 20 trainable parameters. This contrasts with our circuit, which employs 117 trainable parameters. When running their approach on quantum hardware, the authors in \cite{arthur:2022} obtained a classification accuracy for the approach in \cite{arthur:2022} drops to  $45.0\pm 5.0\%$ for the VQC and $28.12\pm 20.62\%$ for the HNN. Its also worth noting that this comparison should be taken with caution since the authors in  \cite{arthur:2022} used the IBM QASM simulator and the IBM Mumbai and IBM Montreal quantum computers, which differ from those used for our results. Moreover, their results are for binary classification, whereas ours are categorical. That is, they employed only the samples corresponding to the iris versicolor and iris virginica classes, whereas we used the three species in the dataset.

\subsection{Regression}

\begin{figure}[t!]
	\centering
	\includegraphics[width=0.8\textwidth, height = 0.8\textwidth]{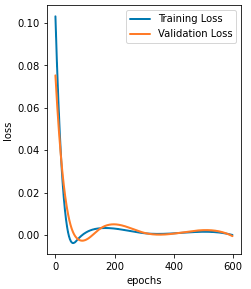}
	
	\caption{Training and validation losses as a function of iteration number for the regression on the Airfoil Self-Noise dataset.}
	\label{fig:airfoil_dataset}
\end{figure}

As mentioned previously, the $\Perthro$ block presented here can also be applied to regression problems in a manner akin to feed forward networks. To illustrate the utility of the $\Perthro$ block for regression, in this section we use the Airfoil Self-Noise dataset \footnote{The Airfoil Self-Noise dataset can be accessed at \url{https://archive.ics.uci.edu/ml/datasets/Airfoil+Self-Noise}}. The structure of the circuit used here is very similar to that used in the previous section. Thus the equivalent network is very similar to the one that applies to our classification experiments. The main difference being that, while the structure of the network follows the same three block-layer pattern, the final block is equivalent to a single neuron layer, i.e. we have substituted the $\Perthro_3$ block in Figure \ref{fig:Pertho_Nets} for a $\Perthro_1$ block. Thus, no additional stages are necessary after the last block. Our regression circuit is thus comprised of blocks three blocks $\Perthro_6$,  $\Perthro_{12}$ and  $\Perthro_1$. We trained our circuit so as to optimise the Mean Squared Error, with and MSE loss which converged after 600 epochs. In Figure \ref{fig:airfoil_dataset}, we show the training and validation loss for our circuit.  % and strong correlation of the outputs to the ground truth results.

% Continue here
When tested on the Armonk, the mean squared error on the testing set is 0.0898. This is as compared to a value of 0.001 on the simulation platform. This is mainly due to the impact of the quantum compounding uncertainty effect in regression problems. It is also worth noting that this effect may be mitigated by using less blocks in the circuit, which would be equivalent to a network with less layers. This is due to the Uncertainty Principle \cite{busch2007heisenberg}, and the consequential fundamental limitation in the accuracy of Quantum Hardware \cite{Giovannetti_2011}, whereby increasing the number of qubits, i.e. layers, would yield more imprecise results.

\section{Conclusion}

In this paper, we have proposed a single-qubit feedforward block whose architecture allows for classical parameters to be used in a manner akin to classical neural networks. To do this, we have departed from the cyclical nature of quantum state probabilities observed on pulsed qubits to show how such a block would be reminiscent of a single-layer neural network comprised of neurons with sine-squared activation functions. In this way, we have shown how pulses can be modulated to excite qubits in order to induce superimposed rotations around the Bloch Sphere. The approach presented here has the advantage of employing a single qubit per block. Thus, being linear concerning the number of blocks rather than polynomial with respect to the number of neurons in a network. Moreover, the approach presented here employs classical parameters and, hence, allows for a large number of training iterations and updates regardless of coherence times. It also has the advantage of allowing for gradients to be reused and stored if necessary. Finally, we have illustrated how the feed forward Pertho block presented here may be used in a manner consistent with neural networks and implemented on an IBM pulse-enabled quantum computer to tackle classification and regression tasks. 

%\section*{Declarations}

%\textbf{Conflict of Interest:} The authors declare no competing interests.
\bibliographystyle{ieeetr}
\bibliography{./MyCollection.bib}

\end{document}